\documentstyle[12pt,preprint,aps,epsf,floats]{revtex}
\def\NPB#1#2#3{Nucl. Phys. {\bf B#1}, #3 (19#2)}
\def\NPBMM#1#2#3{Nucl. Phys. {\bf B#1}, #3 (20#2)}

\def\PLB#1#2#3{Phys. Lett. {\bf B#1}, #3 (19#2)}

\def\PRD#1#2#3{Phys. Rev. {\bf D#1}, #3 (19#2)}
\def\PRDMM#1#2#3{Phys. Rev. {\bf D#1}, #3 (20#2)}
\def\PRL#1#2#3{Phys. Rev. Lett. {\bf#1}, #3 (19#2)}

\def\PRep#1#2#3{Phys. Rept. {\bf#1}, #3 (19#2)}

\newcommand{\mplanck}{M_{Planck}}

\tighten
\begin{document}
\preprint{
\noindent
\begin{minipage}[t]{3in}
\begin{flushleft}
October 2000 \\
\end{flushleft}
\end{minipage}
\hfill
\begin{minipage}[t]{3in}
\begin{flushright}
MIT--CTP--3031\\
CALT--68--2299\\
hep-ph/0010113\\
\vspace*{.7in}
\end{flushright}
\end{minipage}
}

\draft

\title{
More Corrections to the Higgs Mass in Supersymmetry
}

\author{Nir Polonsky$^{a}$ and Shufang Su$^{b}$ 
\vspace*{.2in}
}
\address{
$^{a}$ Massachusetts Institute of Technology, Cambridge, MA 02139}
\address{
$^{b}$ California Institute of Technology, Pasadena, CA, 91125
\vspace*{.2in}}


\maketitle

\begin{abstract}
In supersymmetry, the Higgs quartic couplings is given by the sum
in quadrature of the weak gauge couplings. This leads to the prediction of
a light Higgs boson, which still holds when considering loop corrections
from soft supersymmetry breaking. However, another source of corrections,
which explicitly depends on the scale
of the mediation of supersymmetrey breaking,  
is from generic hard breaking terms. 
We show that these corrections can significantly
modify the Higgs mass prediction in models of low-energy supersymmetry 
breaking, for example, gauge mediation. 
Conversely, the Higgs mass measurement can be
used to constrain the scale of mediation of supersymmetry breaking.
\end{abstract}


Weak-scale supersymmetry provides a well motivated framework, 
both theoretically and phnomenologically, for extending the Standard Model 
of electroweak and strong interactions (SM). Consequently, supersymmetry  
has been the driving force in high-energy research for more than a decade.
Its main attraction is the natural existence 
in supersymmetry of a fundamental Higgs boson.
Furthermore, supersymmetry dictates 
that the quartic coupling in the SM Higgs potential,
\begin{equation}
V = -\mu^{2}hh^{\dagger} + \frac{1}{2}\lambda|hh^{\dagger}|^{2},
\label{VSM}
\end{equation}
is given by a sum in quadrature of the electroweak hypercharge 
and $SU(2)$ couplings, $g\prime$ and $g$, respectively,
\begin{equation}
\lambda = \frac{g\prime^{2} + g^{2}}{4}\cos^{2}2\beta.
\label{lambda}
\end{equation}
The angle $\beta$,
$\tan\beta \equiv\langle H_{2} \rangle / \langle H_{1} \rangle$,
is a parameter of the  
(type II) two-Higgs-doublet model (2HDM)
which generalizes the SM in its supersymmetric extension
with $H_{1}$ ($H_{2}$) coupling to the $b$- ($t$-) quark. 
(The 2HDM is required by anomaly-cancellation constraints and holomorphicity). 
Here, we work in the decoupling limit of the 2HDM in which
one physical Higgs doublet $H$ is sufficiently heavy and
decouples from electroweak symmetry breaking 
while a second SM-like Higgs doublet is 
roughly given 
(up to ${\cal{O}}(M_{Z}^{2}/M_{H}^{2})$ corrections and a phase) 
by $h \simeq H_{1}\cos\beta + H_{2}\sin\beta$,
and we conveniently defined
\begin{equation}
H_n=\left(
\begin{array}{c}
H_n^+\\(H_n^0+iA_n^0)/\sqrt{2}
\end{array}
\right).
\end{equation}
Hence, the mass of the physical SM-like Higgs boson 
$m_{h^{0}}^{2} = \lambda v^{2}$ with $v = \langle h \rangle$
is now bounded at tree level,
\begin{equation}
m_{h^{0}}^{2} \leq M_{Z}^{2}\cos^{2}2\beta,
\label{bound}
\end{equation}
by the $Z$-boson mass, $M_{Z}^{2} = (1/4)
(g\prime^2 + g^{2})v^{2}$. 
This is all described in Ref.~\cite{Hunter}.
Note that $h^{0}$ parameterizes a flat direction of the 2HDM
($\beta \rightarrow \pi/4$), further suppressing its mass
in the appropriate limit.

Supersymmetry, however, must be explicitly and softly
broken with mass splitting
between a fermion $f$ and its scalar superpartner $\tilde{f}$, for example.
Hence, new corrections to the quartic 
coupling arise quantum mechanically, the most important of which, 
\begin{equation}
\delta\lambda_{soft} = 
\frac{3}{8\pi^{2}}y_{t}^{4}\ln\frac{m_{\tilde{t}}^{2}}
{m_{t}^{2}},
\label{deltalambda}
\end{equation}
arises from loops involving the top quark $t$ and its superpartner
the stop $\tilde{t}$, leading to
${\cal{O}}\left((m_{t}^{4}/M_{Z}^{2})\ln({m_{\tilde{t}}^{2}}/
{m_{t}^{2}})\right) \sim {\cal{O}}(100\%)$ radiative corrections to 
$m^{2}_{h^{0}}$ \cite{loop}. The importance of the corrections stems
from the large coupling in the loop, the top-Yukawa coupling
$y_{t} \sim 1$, and from the smallness of the tree-level mass 
Eq.~(\ref{bound}).
(The correction is maximized in the case of large $\tilde{t}_{L} 
- \tilde{t}_{R}$ mixing \cite{LR}, which contributes additional
terms to (\ref{deltalambda}).) 
The upper bound (\ref{bound}) is now corrected by roughly 
a factor of a $\sqrt{2}$
\begin{equation}
m_{h^{0}} \leq 91\, {\rm GeV} \rightarrow
m_{h^{0}} \lesssim 130\, {\rm GeV},
\label{bound2}
\end{equation}
with appropriately much smaller two-loop corrections.
Hence, a strict upper bound on the Higgs boson mass exists
even when soft supersymmetry breaking (SSB) effects are included, 
providing a strong prediction of supersymmetric extensions 
with minimal matter and gauge 
symmetries, commonly referred to hereafter as the MSSM.
(More refined calculations 
can be done in particular MSSM realizations. 
For example, see Ref.~\cite{ssu}.)

 In non-minimal extensions involving, for example, an additional Abelian 
factor in the gauge group ${\rm SM} \rightarrow {\rm SM}\times U(1)$
or a SM singlet coupling to the two
Higgs doublets $y_{s}SH_{1}H_{2}$, 
new tree-level contributions to $\lambda$ appear
(from gauge $D$- and Yukawa $F$-terms, respectively), raising the upper bound 
(\ref{bound}) and modifying its $\beta$-dependence. This provides an important
tool  which discriminates
between minimal and non-minimal realizations of supersymmetry.
However, as long as perturbativity 
is assumed up to Planckian scales, one still has 
$m_{h^{0}} \lesssim 180 - 200\, {\rm GeV}$ \cite{upperbound}. 
(One should bare in mind, however, that these limits are
sensitive to the location of
Landau poles which  appear
in the renormalization of many such models, 
and therefore do not provide as a strict limit 
as the MSSM limit Eq.~(\ref{bound2}).
For examples, see Ref.~\cite{landaupole}.)
Therefore, while the  Higgs mass
may discriminate between models (particularly if $\tan\beta$ is known
independently), its lightness is largely model independent.
A useful but rough approximation of the
Higgs mass is as follows: Each contribution 
to the Higgs mass is at most $\sim 100\%$ of (\ref{bound}) $\sim M_{Z}$, 
and all contributions are summed in quadrature. For example, in models 
with both an extra $U(1)$ and additional SM singlet fields on finds,
including loop corrections, $m_{h^{0}} \lesssim \sqrt{4}M_{Z} \sim 180\, 
{\rm GeV}$ \cite{penn}.

The SSB parameters such as $m_{\tilde{t}}^{2}$ above, 
carry  mass dimensions and can contribute
only logarithmically to the quartic couplings,
and consequently, to the Higgs mass 
(once the vacuum expectation value (VEV) is fixed
to its measured value). The Higgs mass is therefore
more sensitive to the top mass (Yukawa coupling) than to the stop mass.
This is the basis
for the strong and model-independent results for the loop-corrected
Higgs mass in the MSSM. Nevertheless, in general
hard supersymmetry breaking (HSB) quartic couplings also arise
(from non-renormalizable operators in the Kahler potential, for example).
Assuming that the SSB
parameters are characterized by a parameter $m_{0} \sim 1\, {\rm TeV}$
(i.e., $m_{\tilde{t}} \sim m_{0}$) then
\begin{equation}
\delta\lambda_{hard} = 
\tilde{\lambda}_{h}\frac{F^{2}}{M^{4}} \simeq
\tilde{\lambda}_{h}(16\pi^{2})^{2n}\left(\frac{m_{0}}{M}\right)^{2},
\label{lambdahard}
\end{equation}
where $M$ is a dynamically determined scale parameterizing the communication
of supersymmetry breaking to the SM sector, which is distinct
from the supersymmetry breaking scale $\sqrt{F} \simeq
(4\pi)^{n}\sqrt{m_{0}M}$.
Such operators were recently discussed in Ref.~\cite{martin,N2}.
The exponent $2n$ 
is the loop order at which the mediation of supersymmetry breaking 
to the (quadratic) scalar potential occurs. 
(Non-perturbative dynmaics
may lead to different relations that can be described instead 
by an effective value of $n$.)
The coupling $\tilde{\lambda}_{h}$ is an unknown dimensionless coupling
(for example, in the Kahler potential). As long as such quartic couplings
are not arbitrary but are related to the source of the SSB parameters
and are therefore described by (\ref{lambdahard}), then 
they do not destabilize the scalar potential
and do not introduce quadratic dependence on the ultra-violet cut-off
scale, which is identified with  $M$. Consider the one-loop contribution
to a generic mass parameter in the scalar potential, which is given at
tree level by the SSB scale $m^{2}|_{\rm tree} = m_{0}^{2}$, 
\begin{equation}
\delta m^{2} \sim \frac{\delta\lambda_{hard}}{16\pi^{2}}M^{2} =
\tilde{\lambda}_{h}(16\pi^{2})^{2n -1}m_{0}^{2} \lesssim m^{2}|_{\rm tree}.
\label{QD}
\end{equation}
Stabilty of the scalar potential only constrains
$\tilde{\lambda}_{h} \lesssim \min{\left((1/16\pi^{2})^{2n-1}, 1\right)}$
(though calculability and predictability are diminished).

Such a hard coupling corrects (\ref{lambda}) and as a result affects
the tree level Higgs mass bound even in the MSSM.
It introduces an explicit dependence of the Higgs mass on the
supersymmetry mediation scale $M$, a dependence which is
avoided in (\ref{deltalambda}).
In the case that supergravity interactions mediate
supersymmetry breaking from some ``hidden'' sector (where supersymmetry
is broken spontaneously) to the SM sector, one has $M = \mplanck$.
The corrections are therefore negligible whether the mediation
occurs at tree level ($n=0$) or loop level ($n \geq 1$) and can be 
ignored for most purposes. (For exceptions, see Refs.~\cite{martin,sugra}.)
In general, however, the scale of supersymmetry breaking is an 
arbitrary parameter and depends on the dynamics that mediate the SSB
parameters.
For example, it was shown recently that in the case of $N=2$ supersymmetry
one expects $M \sim 1\,{\rm TeV}$ \cite{N2}. 
Also, in models with extra large dimensions
the fundamental $\mplanck$ scale can be as low as  a few TeV,
leading again to $M \sim 1\,{\rm TeV}$.
(For example, see Ref.~\cite{pomarol}.) A ``TeV-type'' mediation scale
implies a similar supesymmetry breaking scale and provides an
unconventional possibility.
(For a discussion, see Ref.~\cite{N2}.) 
It may be further motivated by 
the observation that if the leading contribution to the cosmological constant
(which vanishes in the supersymmetry limit) is $\sim M^{8}/\mplanck^{4}$
then observations suggest that $M \sim 1\,{\rm TeV}$. (Other cosmological
motivations for a TeV dynamical scale were discussed recently
in Ref.~\cite{murayama}.) If indeed $M \sim 1\,{\rm TeV}$ then
$\delta\lambda_{hard}$ given in (\ref{lambdahard}) is ${\cal{O}}(1)$
(assuming tree-level mediation (TLM) and
${\cal{O}}(1)$ couplings $\tilde\lambda_{h}$ in the Kahler potential).
The effects on the Higgs mass must be considered in this case.

A more familiar and surprising example is given by
the (low-energy) gauge mediation (GM) framework \cite{GM}. In GM, 
SM gauge loops communicate between
the SM fields and some messenger sectors, mediating the SSB potential.
The Higgs sector and the related operators, however, are poorly
understood in this framework \cite{mu} and therefore 
all allowed operators should be considered.
In its minimal incarnation (MGM) $2n = 2$, and $M \sim
16\pi^{2}m_{0} \sim 100\, {\rm TeV}$ parameterizes both the mediation 
and supersymmetry breaking scales. 
The constraint (\ref{QD}) corresponds
to $\delta\lambda_{hard} \sim \tilde{\lambda}_{h} \lesssim 1/16\pi^{2}$
and the contribution of $\delta{\lambda_{hard}}$
to the Higgs mass could be comparable to the contribution of
the supersymmetric coupling (\ref{lambda}).
A particularly interesting case is that of non-perturbative messenger dynamics
(NPGM) in which case $n_{eff} = 1/2$,
$M \sim 4\pi m_{0} \sim 10\, {\rm TeV}$ \cite{4pi}, 
and the constraint 
on $\tilde{\lambda}_{h}$ is relaxed to 1. Now
$\delta\lambda_{hard}\lesssim 1$ terms could
dominate the Higgs mass.

The different possibilities are summarized in the 
Table \ref{table:frameworks}.
\begin{table}
\caption{Frameworks for estimating $\delta\lambda_{hard}$.
(Saturation of the lower bound on $M$ is assumed.)}
\label{table:frameworks}
\begin{tabular}{l|l|l|l|l}
&$n$&$\tilde{\lambda}_{h}$& $M$ & $\delta{\lambda}_{hard}$  \\ \hline
TLM &$0$&$\sim 1$& $\gtrsim m_{0}$  & $(m_{0}/M)^{2} \sim 1$ \\
NPGM & $1/2$ & $\sim 1$& $\gtrsim 4\pi m_{0}$  & $(4\pi m_{0}/M)^{2} \sim 1$ \\
MGM  &  $1$ & $\lesssim 1/16\pi^{2}$ & $\gtrsim 16\pi^{2}m_{0}$ &
$(4\pi m_{0}/M)^{2} \sim 1/16\pi^{2}$ 
\end{tabular}
\end{table}
Next, we consider the $\beta$-dependence of the HSB contributions,
which is different from that of all other terms.

In order to address the $\beta$-dependence, 
we revert temporarily to the 2HDM formalism 
of Haber and Hempfling \cite{2hdm}.
The Higgs scalar potential can be 
written down as 
\begin{eqnarray}
V&=&m_{11}^2H_1^{\dagger}H_1+m_{22}^2H_2^{\dagger}H_2-
[m_{12}^2H_1^{\dagger}H_2+{\rm h.c.}] \nonumber \\
&&+\case{1}{2}\lambda_1(H_1^{\dagger}H_1)^2
+\case{1}{2}\lambda_2(H_2^{\dagger}H_2)^2
+\lambda_3(H_1^{\dagger}H_1)(H_2^{\dagger}H_2)
+\lambda_4(H_1^{\dagger}H_2)(H_2^{\dagger}H_1) \nonumber \\
&&+\left\{\case{1}{2}\lambda_5(H_1^{\dagger}H_2)^2+[
\lambda_6(H_1^{\dagger}H_1)+
\lambda_7(H_2^{\dagger}H_2)]H_1^{\dagger}H_2+{\rm h.c.}\right\}.
\label{V}
\end{eqnarray}
Non-zero VEVs for the Higgs fields are obtained at the minimum of the 
scalar potential if $m_{11}^{2}m_{22}^{2} < |m_{12}|^{4} $.
After electroweak symmetry breaking the  five physical 
Higgs degrees of freedom
 (after diagonalizing the mass matrices) are two CP-even bosons
$H^0$ and $h^0$, with $m_{H^0}>m_{h^0}$, one CP-odd scalar $A^0$ and a 
charged Higgs $H^{\pm}$.  
The decoupling limit considered here is defined as $M_{A^0}\gg{M}_Z$:
The heavy physical Higgs doublet $(H^{+}, (H^{0} + iA^{0})/\sqrt{2})^{T}$ 
decouples and the effective theory 
simply reduces to the 
Standard Model with one ``light'' physical Higgs boson $h^{0}$,
$m_{h^0}^2=\lambda{v}^2$.
The effective quartic coupling $\lambda$ 
is related to the quartic couplings $\lambda_{1\ldots{7}}$ 
in the full 2HDM potential (\ref{V}) via 
\begin{equation}
\lambda=c_{\beta}^4\lambda_1+s_{\beta}^4\lambda_2+
2s_{\beta}^2c_{\beta}^2(\lambda_3+\lambda_4+\lambda_5)+
4c_{\beta}^3s_{\beta}\lambda_6+4c_{\beta}s_{\beta}^3\lambda_7,
\label{eq:lambda}
\end{equation}
where $s_{\beta} \equiv \sin\beta$ and $c_{\beta} \equiv \cos\beta$.
Allowing additional hard supersymmetry breaking quartic terms besides the 
usual gauge ($D$-)terms and loop contributions, 
$\lambda_{1\ldots{7}}$ can be written out 
explicitely as 
\begin{eqnarray}
\lambda_{1,2}&=&\case{1}{2}(g\prime^2+g^2)+\delta\lambda_{soft\,1,2}
+\delta\lambda_{hard\,1,2},\\
\lambda_3&=&-\case{1}{4}(g\prime^2 - g^2)+\delta\lambda_{soft\, 3}
+\delta\lambda_{hard\, 3}, \\
\lambda_4&=&-\case{1}{2}g^2+\delta\lambda_{soft\, 4}
+\delta\lambda_{hard\, 4}, \\
\lambda_{5,6,7}&=& \delta\lambda_{soft\, 5,6,7}
+\delta\lambda_{hard\, 5,6,7}.
\end{eqnarray}
The leading SSB contribution $\delta\lambda_{soft\,i}$ were already summed
in (\ref{deltalambda}). The effect of the
HSB contributions $\delta\lambda_{hard\,i}$ 
will be  estimated below.

Substituting all the $\lambda$'s
into Eq.~(\ref{eq:lambda}),
the squared Higgs mass $m_{h^0}^{2}$ reads 
\begin{eqnarray}
m_{h^0}^2&=&\case{1}{4}(g\prime^2+g^2)v^2
\cos^{2}2\beta+\delta{m}^{2}_{loop}
\nonumber \\
&&+
(c_{\beta}^4\delta\lambda_{hard\, 1}+s_{\beta}^4\delta\lambda_{hard\, 2}+
2s_{\beta}^2c_{\beta}^2(\delta\lambda_{hard\, 3}+
\delta\lambda_{hard\, 4}+\delta\lambda_{hard\,5})
\nonumber  \\
&& +
4c_{\beta}^3s_{\beta}\delta\lambda_{hard \,6}+
4c_{\beta}s_{\beta}^3\delta\lambda_{hard \,7})v^2
\end{eqnarray}
\begin{equation}
\hspace{-1 in}{\stackrel{\delta\lambda_{hard\, 1}=\ldots=
\delta\lambda_{hard \,7}=\delta\lambda_{hard}}{=}}
\,\,\, M_Z^2\cos^{2}{2\beta}+\delta{m}^{2}_{loop}+(c_{\beta}+s_{\beta})^4v^2
\delta\lambda_{hard}.
\label{mhall}
\end{equation}
In (\ref{mhall}), we have used the $Z$-boson mass and assumed
for simplicity that all the $\delta\lambda_{hard}$'s are equal.
Note that since no new particles or gauge interactions were introduced,
(\ref{mhall}) reduces to the familiar MSSM result (with only SSB terms)
for $\delta\lambda_{hard\, 1\ldots{7}}=0$.

%
\begin{figure}[t]
\begin{center}
\epsfxsize= 6.5 cm
\leavevmode
\epsfbox[40 150 520 600]{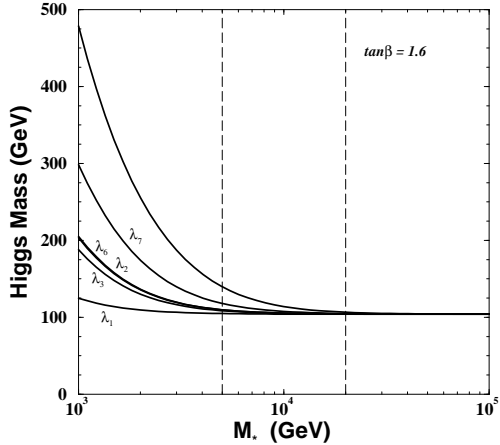}
\end{center}
\caption[f1]{The SM-like Higgs boson mass
for $\tan\beta = 1.6$  is shown
as a function of the effective mediation scale $M_{*}$. 
The contribution of each individual $\delta\lambda_{hard\, i}$
HSB coupling (when added to the result with only SSB, as explained in the text)
is shown (with the label $\lambda_{i}$), 
as well as their sum (upper curve, assuming
that all HSB couplings are equal). 
The $M_{*}$ range in the MGM
case is indicated (dashed lines) for reference. 
For sufficiently large
values of $M_{*}$ all curves converge to the   
MSSM with only SSB upper-bound value.} 
\label{fig:beta1.6}
\end{figure}
\begin{figure}[t]
\begin{center}
\epsfxsize= 6.5 cm
\leavevmode
\epsfbox[40 150 520 600]{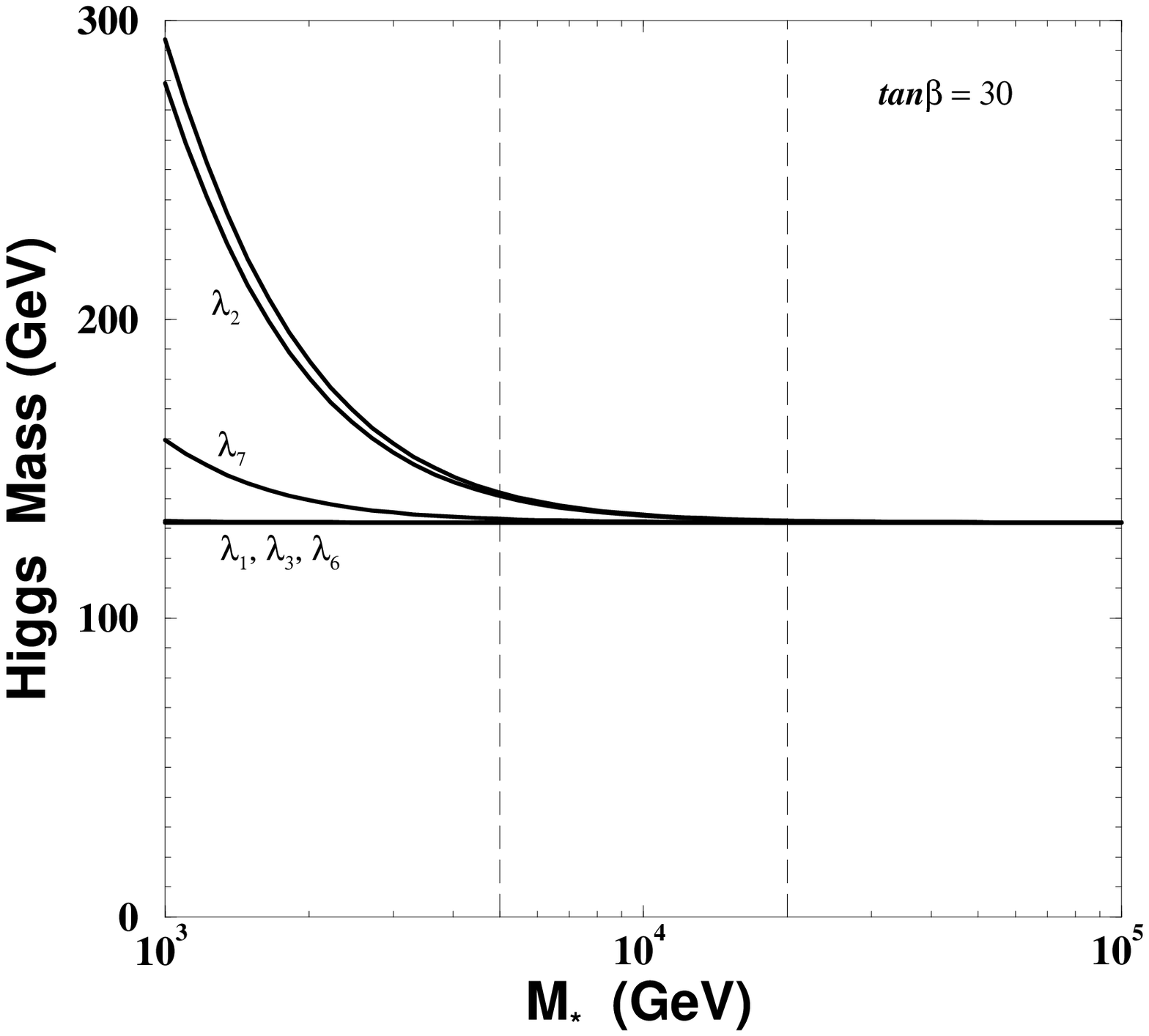}
\end{center}
\caption[f1]{The same as in Fig.~\ref{fig:beta1.6} except
for  $\tan\beta = 30$.}
\label{fig:beta30}
\end{figure}

We are now in position 
to evaluate the HSB contributions to the Higgs mass for
an arbitrary $M$ (and $n$). 
In Figs.~\ref{fig:beta1.6} and \ref{fig:beta30}, we show the dependence 
of the mass of the SM-like Higgs boson on the effective 
mediation scale $M_{*}$, 
including both the loop corrections
(upper-bound values 
adopted from Ref.~\cite{carena} and incorporated numerically) 
and the HSB contributions, for $\tan\beta=1.6$ and $30$ respectively.
The effective scale is defined as
$M_{*} \equiv (M/(4\pi)^{2n}\sqrt{\tilde{\lambda}_{h}})
({\rm TeV} /m_{0})$.
The HSB contributions decouple  
for $M_{*} \gg  m_{0}$, and the results reduce to the MSSM limit
with only SSB (e.g., supergravity mediation).
However, for smaller values of $M_{*}$ the Higgs mass is 
dramatically enhanced.  For $M$=1 TeV and TLM or 
$M$=$4\pi$ TeV and NPGM, both of which correspond to $M_{*} \simeq 1$ TeV, 
the Higgs mass  could be as heavy as  475 GeV for 
$\tan\beta=1.6$ and 290 GeV for $\tan\beta=30$.
For small values of $\tan\beta$, 
the $\delta\lambda_{hard\,7}$ term gives 
the dominant contribution (since 
it is enhanced by a factor of four),
while for large $\tan\beta$ the 
$\delta\lambda_{hard\,2}$ term dominates.

In the MGM case 
$\tilde\lambda_{hard} \lesssim 1/16\pi^{2}$ so that
$M_{*}\sim 4\pi\, {\rm TeV}$
(unlike the NPGM where $M_{*} \sim 1$ TeV).
Given the many uncertainties (e.g., the messenger quantum numbers
and multiplicity and $\sqrt{F}/M$ \cite{GM}) we identify the MGM with
a $M_{*}$-range which corresponds to a factor of two
uncertainty in the hard coupling.
HSB effects are now more moderate but can increase the Higgs mass by 40 (10) 
GeV for $\tan\beta=1.6\ (30)$ (in comparison to the MSSM with only SSB.)
Although the increase in the Higgs mass in this case
is not as large as in the TLM and NPGM
cases, it is of the same order of magnitude as 
or larger than the MSSM two-loop 
corrections \cite{carena}, 
setting the uncertainty range on any such calculation.
Clearly, within the MSSM the Higgs mass could discriminate 
between the MGM and NPDM and help to better
understand the origin of the supersymmetry breaking. 

%
\begin{figure}[t]
\begin{center}
\epsfxsize= 6.5 cm
\leavevmode
\epsfbox[40 150 520 600]{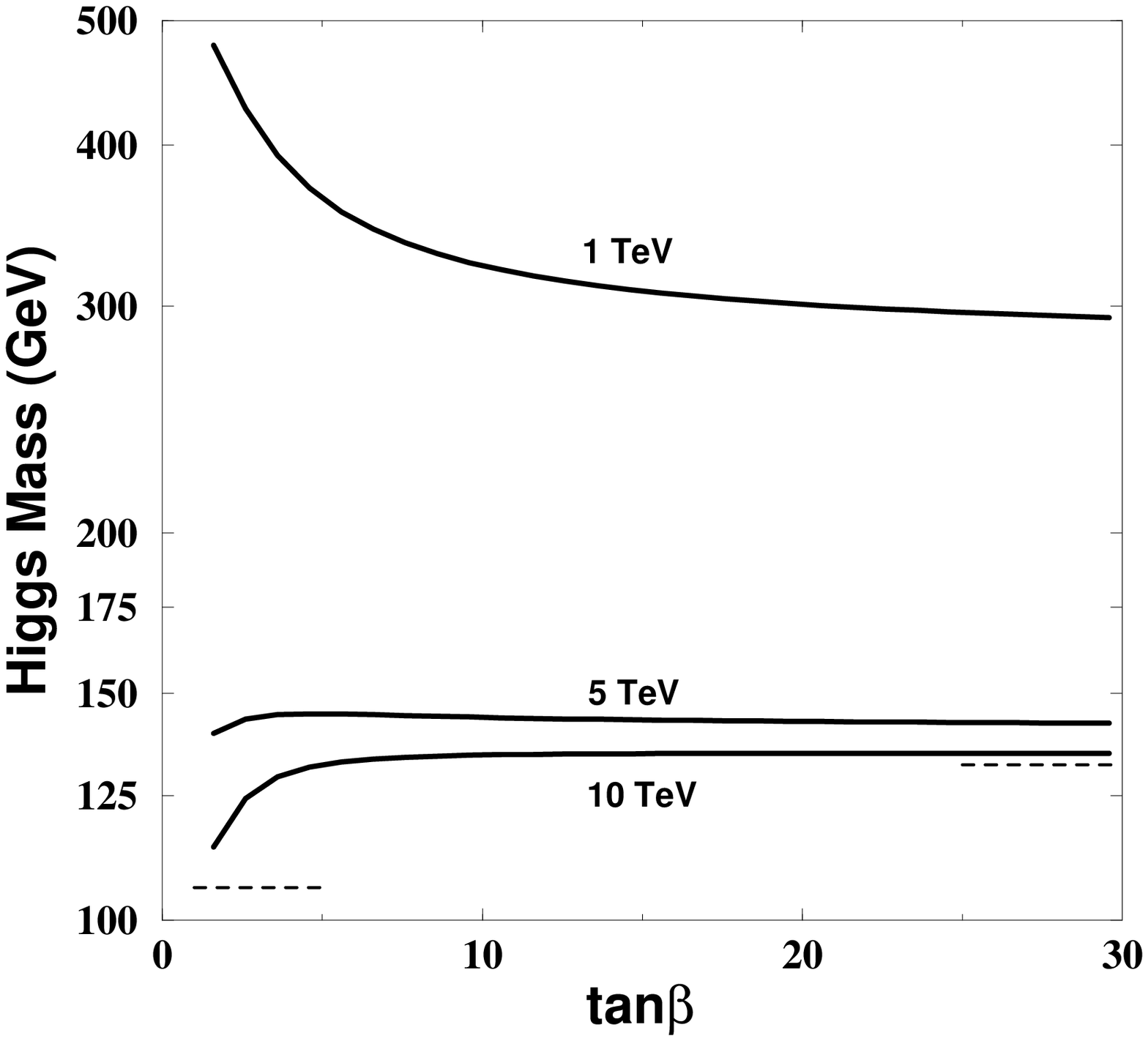}
\end{center}
\caption[f1]{The light Higgs boson mass (note the logarithmic scale)
is shown as a function of $\tan\beta$ for $M_{*}= 1,\,5,\,10$ TeV
(assuming equal HSB couplings).
The upper bound when considering only SSB ($M_{*} \rightarrow \infty$)
is indicated for comparison (dashed lines)
for $\tan\beta = 1.6$ (left) and $30$ (right).}
\label{fig:m1510}
\end{figure}

In Fig.~\ref{fig:m1510}, $m_{h^{0}}$ dependence on $\tan\beta$ for fixed 
values of $M_{*}$
is shown.  The $\tan\beta$ dependence is from
the tree-level mass and from the HSB corrections, 
while the loop corrections to $m_{h^{0}}^2$ are fixed, for simplicity,
at $9200\, {\rm GeV}^2$ \cite{carena}.
The upper curve effectively corresponds to $\delta\lambda_{hard} \simeq 1$.
The HSB contribution dominates the Higgs mass and 
$m_{h^{0}}$ decreases with increasing $\tan\beta$.  As shown above,
$m_{h^{0}}$ could be in the range of $300\ -\ 500$ GeV,
dramatically departing from all previous MSSM calculations
which ignored HSB terms even in the case of low-energy supersymmetry breaking. 
The lower two curves illustrate the range of the corrections in the MGM, 
where the tree-level 
and the HSB contributions compete.  
The $\cos2\beta$ dependence of the tree-level term
dominates the $\beta$-dependence of these two curves.

Following the Higgs boson discovery, 
it should be possible to extract information on the mediation scale $M$.
In fact, some limits can already be extracted.
Consider the upper bound on the Higgs mass derived from a fit to electroweak 
precision data: $m_h^{0}\ <\ 215$ GeV at 95$\%$ confidence level\cite{pdg}.
(Such fits are valid in the decoupling limit discussed here.) 
A lower bound on the scale $M$ in GM could be obtained from 
\begin{equation}
m_Z^2\cos^22\beta+\delta{m}^2_{loop}+(c_{\beta}
+s_{\beta})^4v^2\left(\frac{4\pi{m}_0}{M}\right)^2\leq(215\ {\rm GeV})^2,
\end{equation}
assuming equal $\delta\lambda_{hard}$'s.  
For $\beta=1.6$, it gives 
$M\geq$ 31 TeV while for $\tan\beta=30$ the lower bound is $M\geq$ 19 TeV.
Once $m_{h^{0}}$ is measured, 
more stringent bounds on $M$ could be set.

In conclusion, we illustrated that the scale of the mediation of
supersymmetry breaking explicitly appears in the MSSM prediction
of the Higgs mass, and with a distinct $\beta$-dependence.
(It would also appear in any expression for the Higgs
mass derived in extended models, which correspond to a straightforward 
generalization of our discussion.) In turn, it could lead in certain cases
to a much heavier MSSM Higgs boson than usually anticipated.
It could also distinguish models, e.g., supergravity mediation from 
other low-energy mediation and 
weakly from strongly interacting messenger sectors.
Given our ignorance of the (Kahler potential and) HSB terms,
such effects can serve for setting the uncertainty on any
Higgs mass calculations and can be used to
qualitatively constrain the scale of mediation of supersymmetry breaking
from the hidden to the SM sector.

\acknowledgements

This Work is supported by 
the US Department of Energy under cooperative research 
agreement No.~DF--FC02--94ER40818 
and under grant No.~DE--FG03--92ER40701.

\end{document}